\documentclass[a4paper,11pt]{article}

\usepackage{jinstpub} 

\usepackage{subfigure,dcolumn}
\usepackage{epstopdf}
\usepackage{ulem}

\usepackage{siunitx}

\usepackage{subfigure,dcolumn}
\usepackage{epstopdf}
\usepackage{blindtext}
\usepackage{siunitx}
\usepackage{graphicx}
\usepackage{subfigure}
\usepackage{bm}
\usepackage{lineno}
\usepackage{amsmath}
\usepackage{overpic}

\def\mkred#1{{\color{red}#1}}

\newcommand{\bcntr}{\begin{center}}
\newcommand{\ecntr}{\end{center}}
\newcommand{\beq}{\begin{equation}}
\newcommand{\eeq}{\end{equation}}
\newcommand{\beqar}{\begin{eqnarray}}
\newcommand{\eeqar}{\end{eqnarray}}
\newcommand{\bitm}{\begin{itemize}}
\newcommand{\benu}{\begin{enumerate}}

\newcommand{\bitmb}{\begin{itemize}}
\newcommand{\benub}{\begin{enumerate}}
\newcommand{\eitm}{\end{itemize}}

\newcommand{\bfrm}{\begin{frame}}
\newcommand{\efrm}{\end{frame}}
\newcommand{\bct}{\begin{center}}
\newcommand{\ect}{\end{center}}
\newcommand{\bclm}{\begin{columns}}
\newcommand{\eclm}{\end{columns}}
\newcommand{\bpic}{\begin{overpic}}
\newcommand{\epic}{\end{overpic}}
\newcommand{\bblk}{\begin{block}}
\newcommand{\eblk}{\end{block}}

\newcommand{\eenu}{\end{enumerate}}

\newcommand{\us}{\si{\micro\second}}

\newcommand{\ps}{\si{\pico\second}}
\newcommand{\s}{\si{\second }}

\newcommand{\nm}{\si{\nano\metre}}

\newcommand{\mm}{\si{\milli\metre}}

\newcommand{\mV}{\si{\milli\volt}}

\newcommand{\ns}{\si{\nano\second}}

\newcommand{\ohm}{\si{\ohm}}

\newcommand{\mhz}{\si{\mega\Hz}}
\newcommand{\ghz}{\si{\giga\Hz}}

\newcommand{\gev}{\hbox{GeV}}

\newcommand{\mev}{\hbox{MeV}}
\newcommand{\kev}{\hbox{keV}}

\newcommand{\baft}{\hbox{BaF}_2}
\newcommand{\bafty}{\hbox{BaF}_2\hbox{:Y}}
\newcommand{\yft}{\hbox{YF}_3}

\def\Journal#1#2#3#4{{#1} {\bf #2} (#4) #3}

\def\NIMA{Nucl. Instrum. Methods A}

\def\FoP{Front. Phys.}
\def\NST{Nucl. Sci. Tech.}
\def\inno{Innovation}

\def\tns{IEEE Trans. Nucl. Sci.}
\def\jacry{J. Appl. Crystallogr}
\def\JINST{JINST}

\title{\boldmath Scintillation and Timing Performance of a 3at\% Yttrium-Doped Barium Fluoride Crystal}

\author[a]{Zeyu Huang,}
\author[b,c, 1]{Jing Zhang \note{First Author and Second Author contributed equally to this work}, }
\author[a]{Shiming Zou,}
\author[a]{Mingkuan Yuan,}
\author[b,c]{Jiawei Xu,}
\author[a]{Xiyang Wang,}
\author[a]{Shiqing Xie,}
\author[a,2]{Jinhui Chen \note{Corresponding author}, }
\author[b,c,3]{Junfeng Chen \note{Corresponding author}, }
\author[a,4]{and Xiaolong Wang \note{Corresponding author.}}

\affiliation[a]{Key Laboratory of Nuclear Physics and Ion-beam Application (MOE) and Institute of Modern Physics, 
Fudan University, \\
220 Handan Road, Shanghai, 200433, China}

\affiliation[b]{Center of Materials Science and Optoelectronics Engineering, University of Chinese Academy of Sciences,
Beijing 100049, China}

\affiliation[c]{R\&D Center for Novel Materials, Shanghai Institute of Ceramics, Chinese Academy of Sciences, Shanghai
201899, China}

\emailAdd{chenjinhui@fudan.edu.cn}
\emailAdd{jfchen@mail.sic.ac.cn}
\emailAdd{xiaolong@fudan.edu.cn}

\abstract{
We report the scintillation and timing performance of a newly developed $200~\mm \times 20~\mm \times 20~\mm$ large size  barium fluoride crystal doped with
3at\% yttrium ($\bafty$) to enhance the application for high time resolution. This doping effectively suppresses the
slow scintillation component while maintaining most of the fast component, as confirmed by X-ray excited luminescence
measurements. The $\bafty$ crystal demonstrated a transmittance of near 90\% in the visible spectrum and a light
response uniformity parameter of $\delta = (-2.74 \pm 1.15)\%$ when coupled with the tail end. The actual yttrium
content varied from 2.1at\% near the seed end to 3.7at\% at the tail end. The assembled large $\bafty$ detector with
silicon photomultipliers exhibited a time resolution of $82.2 \pm 2.6~\ps$ using constant fraction discrimination method
in a cosmic ray test and $140.1 \pm 3.8~\ps$ using a low fixed threshold method in a beam test at Shanghai Synchrotron
Radiation Facility with a $1.35~\gev$ electron beam. These results indicate the significant potential of $\bafty$
crystal for various applications, such as detectors for particle physics and nuclear physics.}

\keywords{Barium fluoride, Yttrium-doped, Ultra-fast detector, Timing, Scintillation}

\begin{document}
\maketitle

\section{Introduction}
\label{sec:intro}

Inorganic scintillators play a crucial role in the development of electromagnetic calorimeters (ECALs) utilized in high-energy physics (HEP) and nuclear physics experiments. Additionally, precise timing is essential for spectrometers, notably the Time-of-Flight (TOF) detector used in relativistic heavy ion collider experiments~\cite{RHIC}. The next generation of HEP experiments—including the Circular Electron Positron Collider, the Super Tau-Charm Facility, and the upgrade of the Belle II experiment—demands ultra-fast scintillators capable of meeting increasingly stringent time resolution requirements \cite{cepc1, cepc2, stcf, B2Upgrade}. The ECAL designs for some projects necessitate scintillators with a
timing capability of about $100~\ps$ and robust radiation tolerance. Meanwhile, ultra-high-frequency
radiation imaging and detection are also critical for probing internal nuclear reaction processes and assessing the
structural integrity of materials. Facilities utilizing XFELs, such as LCLS, SACLA, and SwissFEL, have
developed ultra-fast imaging technologies that significantly enhance their experimental timing
capabilities~\cite{ref7,ref8}.

Among various scintillating materials, barium fluoride ($\baft$) has emerged as a promising candidate due to its
remarkable properties. Characterized by an ultra-fast cross-luminescence component with a decay time of approximately
$0.6~\ns$, $\baft$ demonstrates superior timing capabilities~\cite{ref9}. $\baft$ also stands out as a cost-effective
scintillator with great potential for mass-producing large crystals. Consequently, $\baft$ has been adopted as the
bottom-line scintillator of choice for Mu2e-II ECAL and is uniquely positioned to fulfill the stringent requirements
for sub-nanosecond time resolution in Project X~\cite{ref10}.

The implementation of TOF technology requires scintillators with an exceptional time resolution of
less than $100~\ps$, a criterion where the ultra-fast inorganic scintillator $\baft$ excels~\cite{ref11}. With
advancements in quantum efficiency of photon detection from silicon photomultipliers (SiPMs), $\baft$ detector can
achieve a time resolution as low as $51 \pm 3~\ps$ with $2~\mm \times 2~\mm \times 3~\mm$ crystals, highlighting its
immense potential for ultra-fast timing applications in TOF systems~\cite{ref12}. However, this time-resolved result
is measured when the size of the crystal is relatively small, and newer detectors often require large size inorganic
scintillator crystals~\cite{lar1}. At the same time, the CMS upgrade plan to solve the problem of signal accumulation
that may occur later, the LYSO:Ce crystal of $3~\mm \times 3~\mm \times 57~\mm$ is selected~\cite{large, cms01}. This
detector will measure tens of picoseconds of time in its barrel, but the crystal size is not large compared to the
organic scintillators used for TOF. There is potential for using large size $\baft$ crystals for timing detectors in
large future HEP facilitates.

However, $\baft$ also presents a slow component featuring emission peaks in the range of $300-310~\nm$ and a decay
time of about $0.6~\us$, which provides a light output that is 4 to 5 times greater than that of the fast 
component~\cite{ref13, ref14}. This dual-component behavior poses significant challenges, particularly in 
high-counting-rate scenarios, as it can lead to substantial signal pile-up and also deteriorates its timing
performance~\cite{ref15, visvikis}. This issue is critical for future intensity frontier HEP experiments using intense
particle beams that require ultra-fast precise radiation measurements. Incorporating $\mathrm{Y^{3+}}$ (yttrium) into
the crystal structure, denoted as $\bafty$, can effectively suppress the slow scintillation component while preserving
the desirable characteristics of the ultra-fast scintillation~\cite{jfchen}. In 2019, to meet the demands for $\ghz$
hard X-ray imaging at the Matter-Radiation Interactions in Extreme (MaRIE) free-electron laser facility,
Chen Hu {\it et al.} proposed the use of $\bafty$ crystals for ultra-fast, full-absorption scintillator-based
imaging~\cite{ref11}.

Using the Bridgman method, we grew and fabricated a \textcolor{red}{$200~\mm \times 20~\mm \times 20~\mm$ large size} $\bafty$ single crystal boule with a doping concentration of 3at\%
(atomic percentage of cationic atoms) yttrium and a length of $250~\mm$. We found that the 3at\% yttrium concentration
is the optimization to suppress the slow component and ensure the fast component's light output. A comprehensive
analysis of this scintillator's optical and pertinent characteristics was conducted, followed by determining its time
resolution within the context of cosmic rays and high-energy electron beams in a TOF setup. In this study, we report
the application of large $\bafty$ for high-energy experiments for particle physics and nuclear physics and measure the related properties first.

\section{Scintillation characteristics of $\bafty$}

\begin{figure}[htbp!]
\centering
\includegraphics[width=0.4\textwidth]{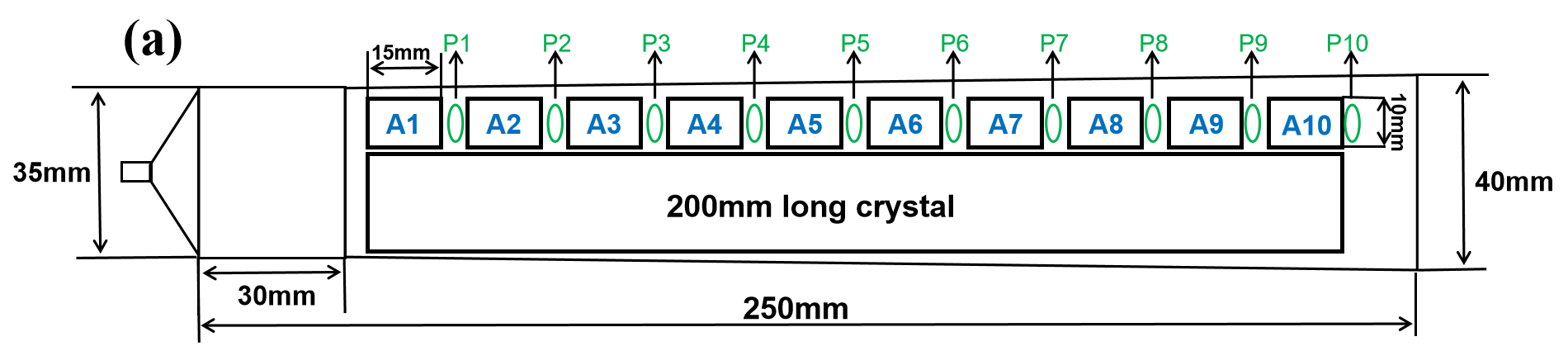}\\
\includegraphics[width=0.4\textwidth]{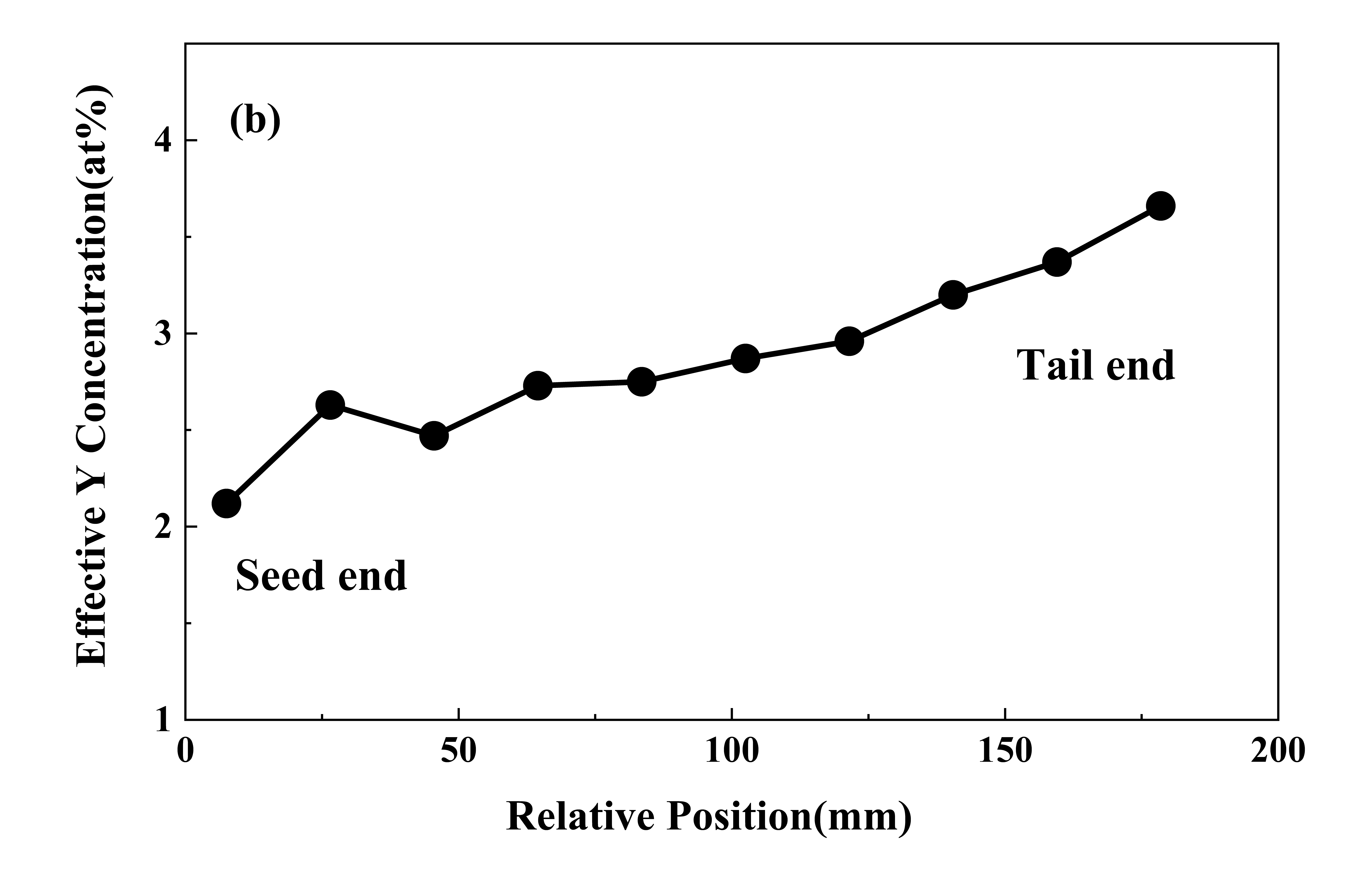}
\caption{(a) The relative positions of the samples in the $\bafty$ ingot. (b) The actual Y doping
concentration at different positions in the crystal.}
\label{position}
\end{figure}

High-purity raw materials, $\baft$(4N) and $\yft$(5N) were accurately weighed according to a molar ratio of $97:3$,
and subsequently homogenized and placed into graphite crucibles for vacuum Bridgman method within a growth rate of
$2~\mm/h$. Figure~\ref{position}(a) shows the relative position of all samples used in the experiments in as-grown
$\bafty$ boule before machining. The samples surrounding the $200~\mm$-long $\bafty$ crystal were divided into equal
segments from the seed end to the tail end. Samples P1-P10 were ground into powder for inductively coupled plasma
atomic emission spectroscopy (ICP-AES) analysis to determine the effective concentration distribution of yttrium
within the crystal. \textcolor{red}{The relative uncertainty of each ICP-AES sample data is about 1\%.} Additionally, samples A1-A10 with a dimension of $15~\mm \times 10~\mm \times 10~\mm$, were cut
and fabricated for X-ray excited luminescence (XEL) testing.

As shown in Fig.~\ref{position}(b), the effective concentration of yttrium increases gradually from 2.12at\% at the
seed end to 3.66at\% at the tail end. The uniformity of doping within the crystal material is typically assessed by
the effective segregation coefficient, $K_{\rm eff}$, which be calculated by solving the equation:
\begin{equation}\label{Keff}
\frac{C_s}{C_o} = K_{\rm eff} \times (1-g)^{K_{\rm eff}-1},
\end{equation}
where $C_{o}$ and $C_{s}$ are the dopant concentrations in the crystal and the melt, respectively, and $g$ is the
solidification fraction of the melt, refers to the volume ratio of the part solidified into the crystal to the whole
ingot during the crystal growth~\cite{FYang}. The $K_{\rm eff}$ of the $\bafty$ crystal was determined to be
$0.75 \pm 0.03$. Compared with La, which has a $K_{\rm eff} \approx 1.53$ in $\baft$ crystal, the measured
$K_{\rm eff}$ here indicates that more uniform yttrium doping has been achieved in $\baft$ crystal.

\begin{figure}[htbp!]
\centering
\includegraphics[height=6cm]{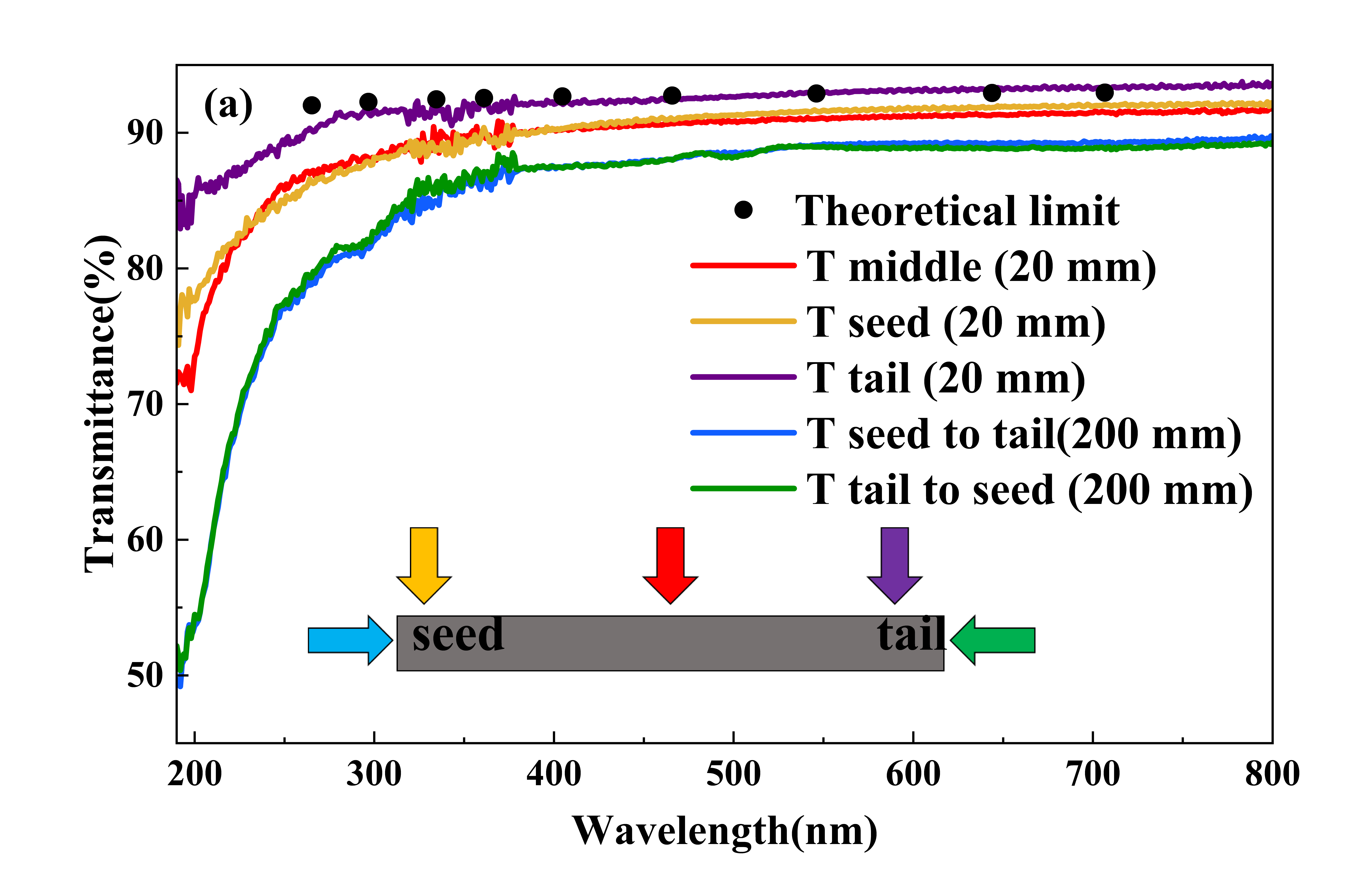}
\includegraphics[height=6cm]{XEL.png}
\caption{(a) The transmittance in different positions of the crystal with different optical paths. \mkred{The theoretical limits of $\bafty$ transmittance are also shown for comparison~\cite{ref11, CHu}.} (b) The XEL curves
for different doping concentrations.}
\label{transm}
\end{figure}

We conducted separate measurements of the transmittance spectra along the lateral direction with a $20~\mm$ optical path at the \textcolor{red}{tail end, seed end, and the midpoint} of the $200~\mm$-long $\baft$:Y crystal and along the growth direction with a $200~\mm$ optical path, within the range of $190~\nm$ to $800~\nm$. Figure~\ref{transm} shows an 80\% lateral transmittance at $200~\nm$ and maintains a transmittance close to 90\% within the visible spectrum. \textcolor{red}{The transverse transmittance at the seed end was slightly lower than at the tail end, which was caused by the unstable growth process. The crystal exhibits yet-to-be-improved transmittance along the growth direction, which was also influenced by the quality difference at the seed end.} Compared to the previous 190 mm-long $\baft$ crystal doped with 1at\% yttrium reported in 2018, the optical quality has improved without significant Ce$^{3+}$ absorption at $290~\nm$~\cite{limitation}. Consequently, there is considerable scope for further enhancement in the optical quality of $\baft$:Y through an optimized growth process.

Different concentrations of $\baft$ crystals exhibit distinct suppression levels of the slow scintillation component.
From the former XEL spectra study, there are two prominent peaks in the range of $200~\nm$ to $450~\nm$, corresponding
to the fast \textcolor{red}{core valence luminescence} component at $220~\nm$ and the slow STE component at $300~\nm$~\cite{ref11, CHu}.
Figure~\ref{transm}(b) presents the results of the XEL spectral analysis conducted on the $\bafty$ crystals, with
measurements taken from the seed end (A1) to the tail end (A10) of the A series samples, in comparison to a pure
$\baft$ sample [UA in Fig.~\ref{transm}(b)]. The original 3at\% yttrium doping significantly reduces the intensity of
the slow component's emission to approximately 20\% of the intensity observed in pure $\baft$ crystals.
Figure~\ref{transm}(b) also emerges a noticeable trend. As the yttrium content increases from 2.12at\% near the seed
end to 3.66 at\% at the tail end, the luminescence intensity of the slow component consistently declines. The
suppression effects on the slow component due to yttrium segregation are slightly more pronounced from the seed end
to the tail end, with variations not exceeding 5\%. Conversely, the luminescence intensity of the fast component
exhibits only marginal changes relative to the intensity of the pure crystal, indicating a selective impact of
yttrium on the crystal's radioluminescence properties.

\begin{figure}[htbp!]
\centering
\includegraphics[width=0.4\textwidth]{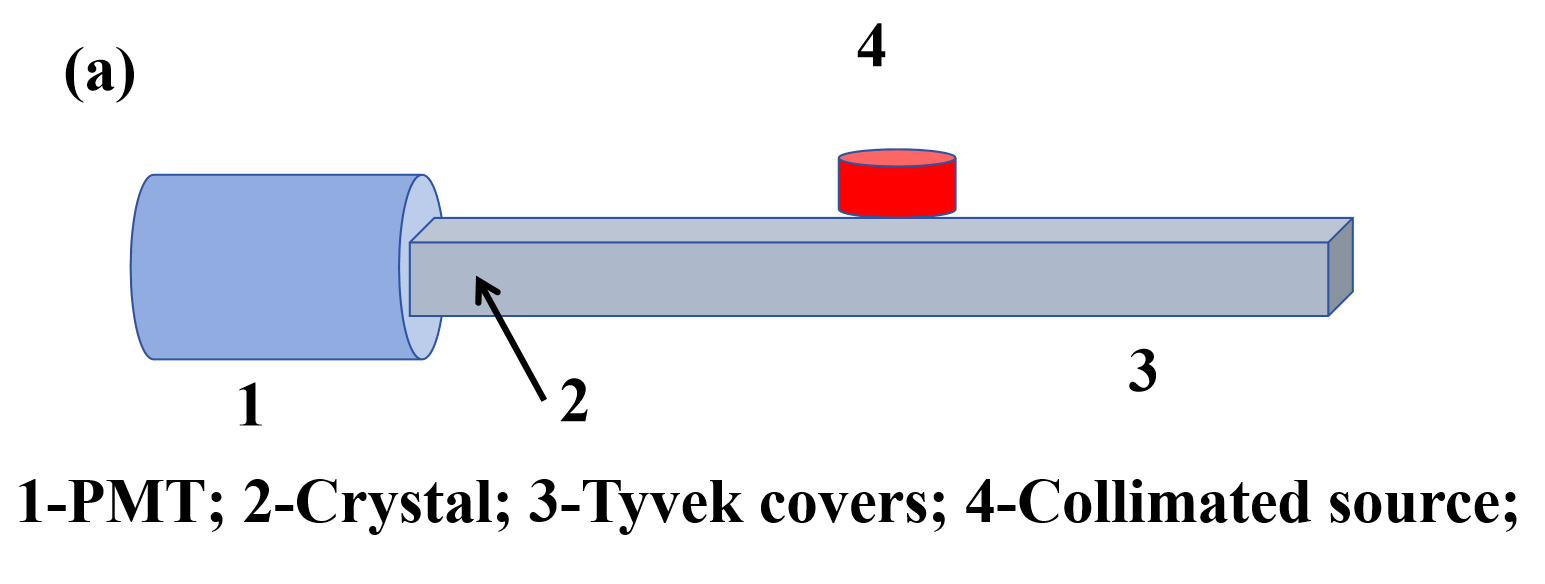}\\
\includegraphics[width=0.4\textwidth]{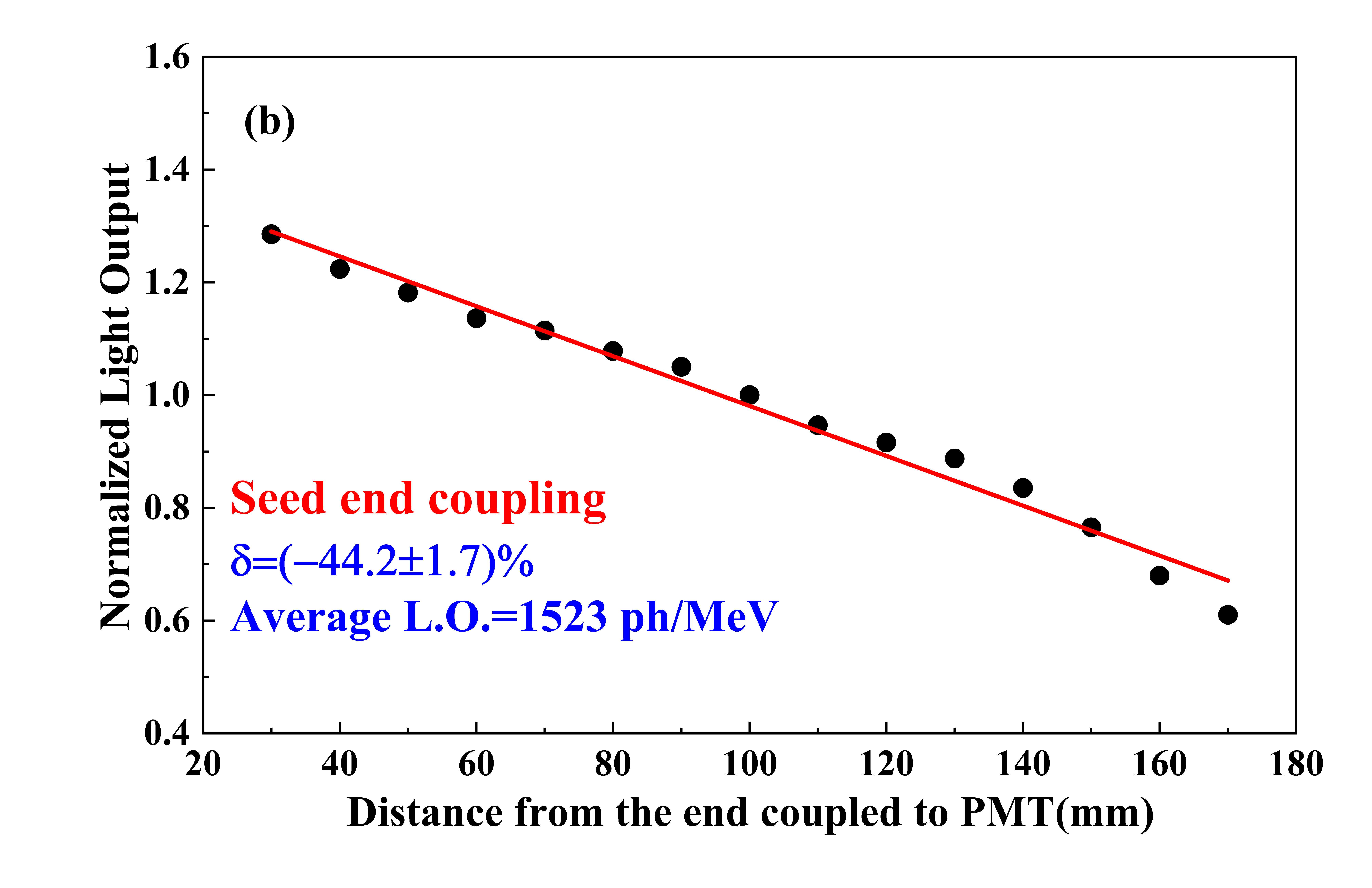}\\
\includegraphics[width=0.4\textwidth]{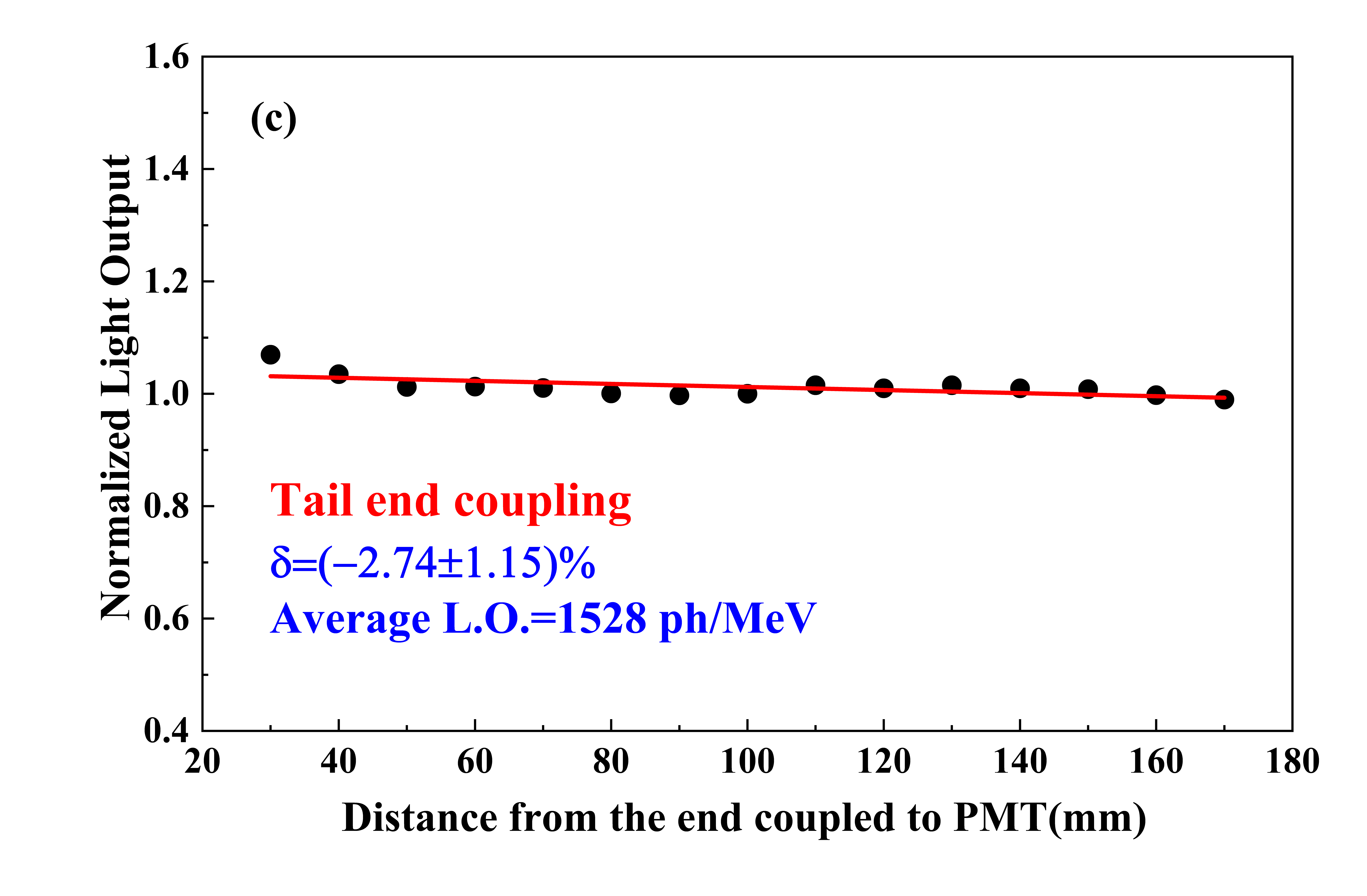}
\caption{(a) Diagram of the LRU experimental setup. (b) and (c) Fitting results from the LRU experiment.}
\label{lre}
\end{figure}

We conducted light response uniformity (LRU) testing of the $\bafty$ crystal using a collimated $^{137}\text{Cs}$
source to explore the relationship between the normalized $3~\us$ gate relative light output and the distance from the
coupling surface to the radiation source~\cite{LYYuan}. This investigation coupled the Hamamatsu R2059 photomultiplier
tube (PMT) at the crystal's seed or tail end to form a single-end readout mode. To enhance the light collection
efficiency, the coupling surface was treated with Dow Corning PMX-200 silicone oil, and the other surface was wrapped
with a single layer of Tyvek. The collimated $^{137}\text{Cs}$ source emitted gamma rays at equidistant points along
the length of the crystal to measure light output at various excitation positions. The single-end readout experimental
setup was established to evaluate the LRU performance of the $\bafty$ crystal when excited by gamma rays, as depicted
in Fig.~\ref{lre}(a). The absolute light output is defined as
\begin{equation}\label{lo}
{\rm LO}(ph/\mev) \equiv \frac{\mathcal{A}}{\mathcal{A}_{\rm SPE}}\frac{1}{E_\gamma}\frac{1}{\epsilon_{\rm EWQE}},
\end{equation}
where $\mathcal{A}$ represents the channel of the full energy peak in pulse height spectra, $\mathcal{A}_{\rm SPE}$ is
the channel of the single photoelectron peak in pulse height spectra under the same test conditions, and $E_\gamma$
denotes the gamma-ray energy from the $^{137}\text{Cs}$ source equal to $662~\kev$. The emission-weighted quantum
efficiency $\epsilon_{\rm EWQE}$ of the $\bafty$ scintillator when coupled with the R2059 PMT was calculated to be
21.6\%.

The deviation parameter for LRU, denoted as $\delta$, is defined by the solving the equation
\begin{equation}\label{fit1}
\frac{{\rm LO}}{{\rm LO}{\rm mid}}=1+\delta\left(\frac{X}{X {\rm mid}}-1\right),
\end{equation}
where $X$ is the source position from the coupling end, and $\rm LO_{mid}$ is the fitted value of the light output at
the midpoint $X_{\rm mid}$ of the crystal. Generally, an absolute value of $\delta$ less than 5\%  is considered
desirable for scintillation crystals used in calorimetry applications.

The fitting results from the LRU experiments, presented in Figs.~\ref{lre}(b) and \ref{lre}(c), reveal that when the
$\bafty$ crystal was coupled at the seed end, the corresponding $\delta$ value was $(-44.2\pm 1.7)\%$. In contrast,
coupling at the tail end resulted in a $\delta$ value of $(-2.74\pm 1.15)\%$. Furthermore, the average light output
measured from the seed and tail ends was found to be $1523~ph/\mev$ and $1528~ph/\mev$, respectively.

Although the average light output of the two ends is nearly equal, the LRU performance shows an obvious difference.
The LRU performance, when coupled from the seed end, is significantly compromised due to defects in the seed region and an increase in yttrium concentration. The poor transmittance leads to increased scattering light loss, adversely affecting overall LRU performance, while the elevated yttrium concentration contributes to a similar negative trend. In contrast, coupling from the tail end exhibited superior LRU performance, with a $\delta$ value within the acceptable range of less than 5\%. This improvement can be attributed to a more favorable balance of light transmission loss and a decrease of light output by increasing the actual yttrium concentration when coupled from the tail end. These findings suggest that for long $\bafty$, coupling from the tail end could 
offer greater potential for applications.

Looking ahead, we anticipate that further advancements in growth conditions and increasing the optical transmittance
of $\bafty$ crystal will improve the LRU of the $\bafty$ crystal, thereby enhancing its suitability for various
experimental applications.

\section{Timing performance of $\bafty$ }

\begin{figure}[htbp!]
\centering
\begin{overpic}[width=0.6\textwidth]{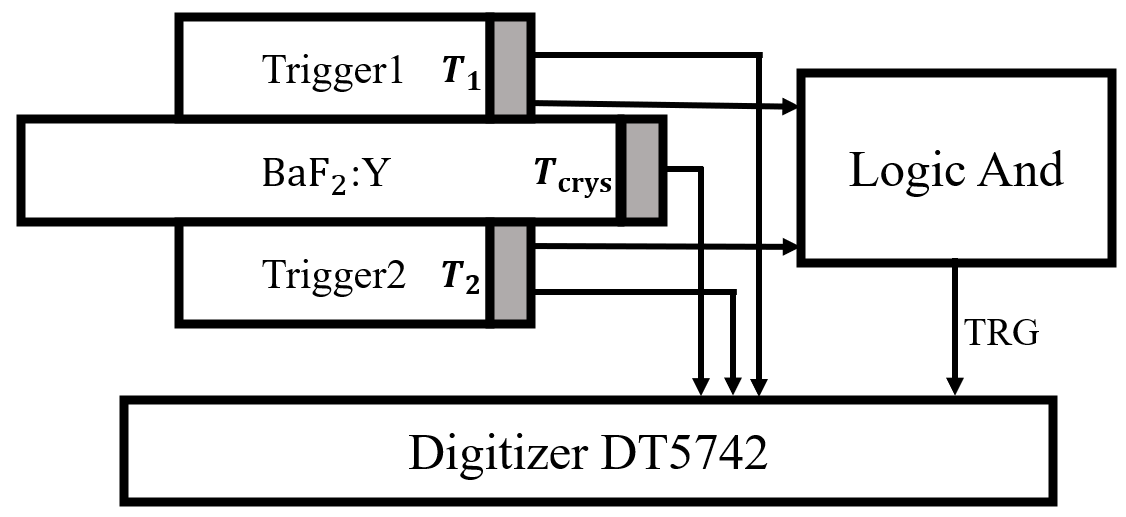}
\put(1,40){ \color{black}{\bf \large (a) }}
\end{overpic}
\begin{overpic}[width=0.6\textwidth]{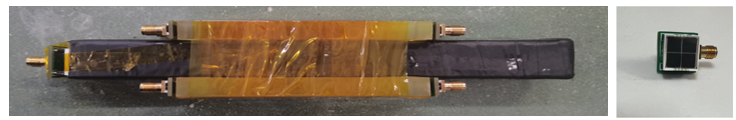}
\put(1,12){ \color{black}{\bf \large (b) }}
\end{overpic}
\caption{Experimental setup for time measurement. (a) Schematic diagram of the setup. (b) Assembled $\bafty$ detector along with the SiPMs.}
\label{setup}
\end{figure}

Cosmic rays and an electron beam in the Shanghai Synchrotron Radiation Facility (SSRF) were utilized to evaluate the
time resolution of a detector prototype built with the $\bafty$ crystal and a SiPM array, thereby simulating practical
scenarios encountered in various applications~\cite{MHJiang,WYCheng,nst03}. Figure~\ref{setup}(a) illustrates the experimental setup employed for measuring time resolution. The trigger system consisted of two plastic scintillators at the top and bottom of the $\bafty$ detector. Waveform acquisition was performed using the CAEN digitizer DT5742, boasting a sampling rate of 5 GS/$\s$ and a bandwidth of $500~\mhz$, delivering
excellent performance in high-speed signal acquisition applications. Initially, the two signals from
the plastic scintillators were coincident, and the logic signals were then
routed into TR0 of the DT5742 to serve as the system's trigger.

Figure~\ref{setup}(b) shows the assembled $\bafty$ detector. All surfaces of the $\bafty$ crystal were thoroughly polished. Except for the face coupled to the SiPM array, the crystal was wrapped in a single layer of Tyvek, followed by a layer of black insulating tape, to enhance light reflection at its surfaces. We selected the fourth-generation VUV-MPPC S13371-6050CQ-02, optimized for photon detection in the ultraviolet region, for the $\bafty$ detector~\cite{Pershing,Vachon}. \mkred{The array has four VUV-MPPCs, each having a sensitive area of $6~\mm\times 6~\mm$.}  The type of SiPM is particularly well-suited for detecting the luminescence emitted by $\bafty$ crystals, especially the fast scintillation component that peaks around $220 \nm$. 
\textcolor{red}{The four VUV-MPPCs were connected in series to achieve a fast rise time, thereby enhancing the time resolution performance of the detector.}
The two plastic scintillators have the same dimension of $40~\mm \times 10~\mm \times 100~\mm$. Each bar was coupled at one end of the cross section ($40~\mm \times 10~ \mm$) to an array of four Hamamatsu S14160-6050 MPPCs using PMX-200 silicone oil. The scintillators were subsequently wrapped in aluminum foil and covered with insulating tape to prevent light leakage.

To get better time resolution, we developed a novel high-speed preamplifier with a bandwidth of $426~\mhz$, a baseline noise of approximately $0.6~\mV$, and a good time resolution better than $20~\ps$~\cite{pream}. 
Upon digitizing the waveform with the digitizer, the resulting data is processed offline on a computer. 
After analyzing the waveforms from DT5742, we extracted comprehensive data regarding the ADC values, Time-to-Digital
Converter measurements, baseline stability, and timing precision~\cite{HYZhang}. To accurately determine the
arrival time of the signals, we fitted the rising edge of the waveform and then employed the constant fraction discrimination (CFD) method.

\subsection{Cosmic ray test}
\label{cosmic ray}

\begin{figure}[htbp!]
\centering
\includegraphics[width=0.4\textwidth]{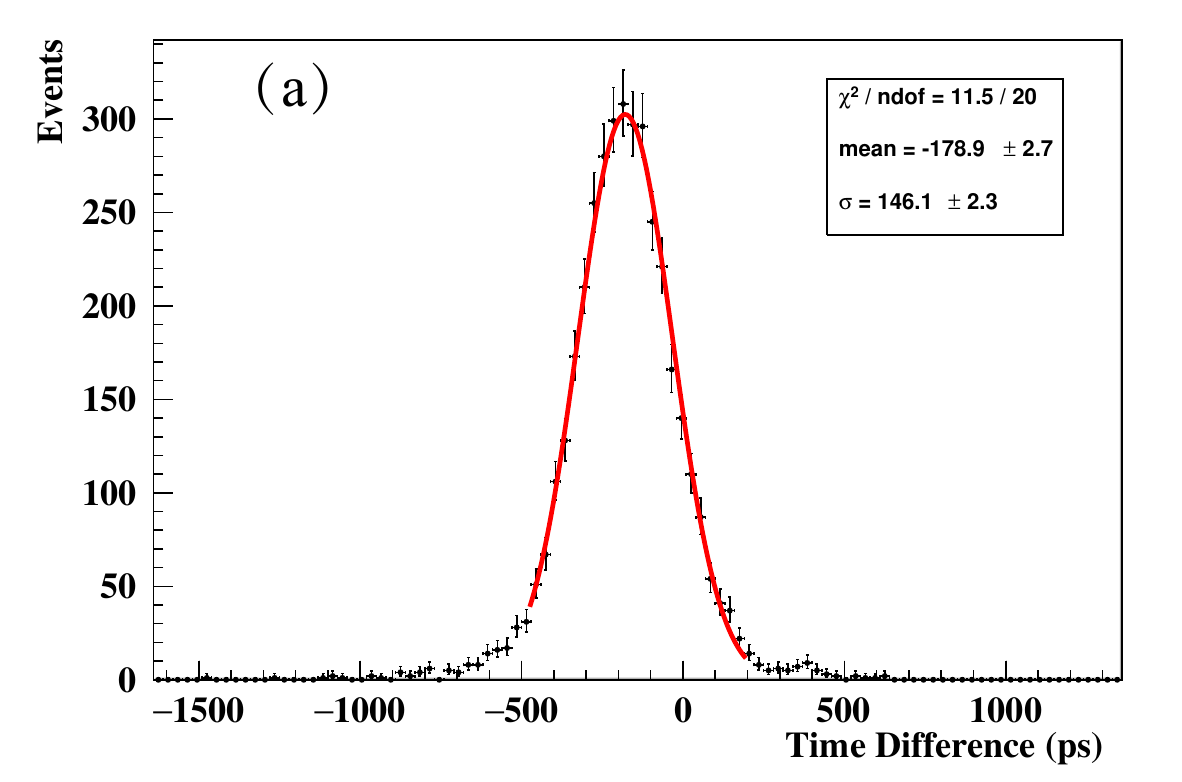}
\includegraphics[width=0.4\textwidth]{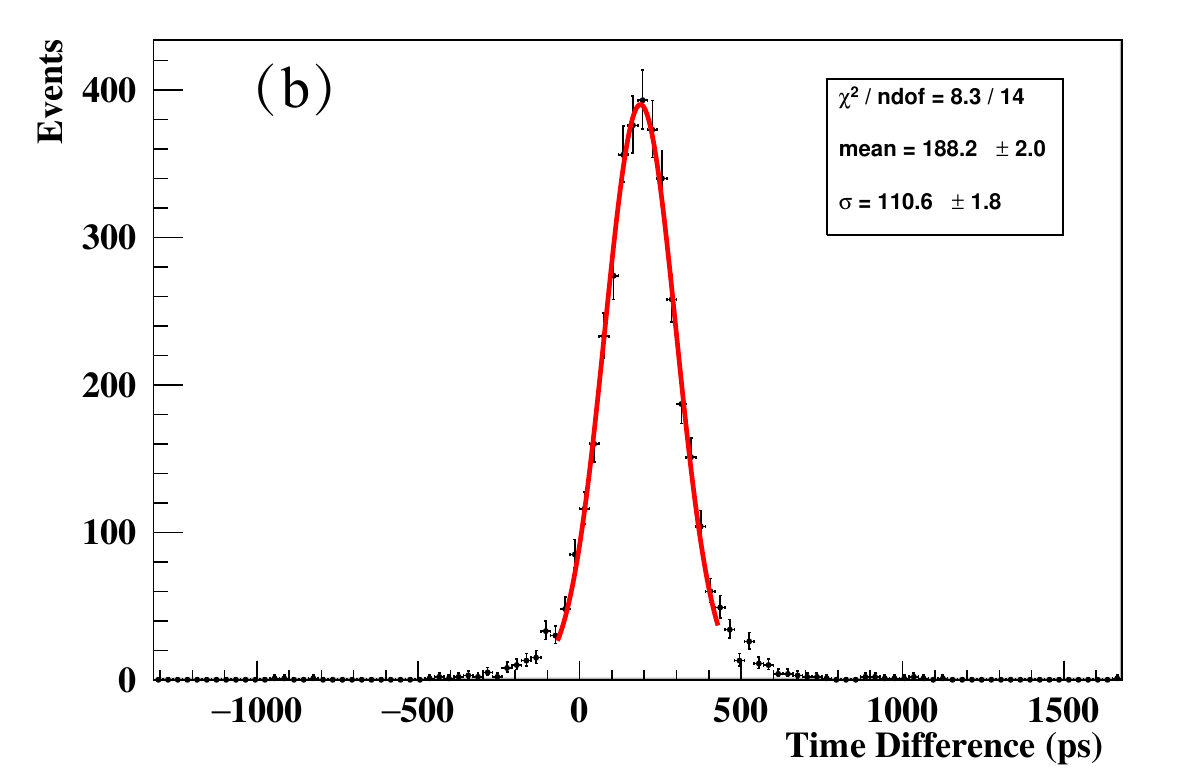}
\caption{Time difference distribution in the cosmic ray test for (a) the trigger system and (b) the entire test system. The red line indicates the Gaussian fit to the data. }
\label{CR_test}
\end{figure}

As shown in Fig~\ref{setup}(a), the trigger time $T_0$ of the test system was calculated by averaging the times recorded by the two plastic scintillators: $T_0 = (T_1 + T_2)/2$, where $T_1$ and $T_2$ are of the two scintillators. The time difference between the two plastic scintillators denoted as $\Delta T_{\rm trg} = T_1 - T_2$, was used to calculate the time resolution of the trigger system, represented as $\sigma(T_0) = \sigma(T_{\rm trg})/2$. The distribution of $\Delta T_{\rm trg} = T_1 - T_2$ is shown in Fig.~\ref{CR_test}(a). By fitting this distribution with a Gaussian function, we obtained a standard deviation of $\sigma(\Delta T_{\rm trg}) = 146.1 \pm 2.3~\ps$, resulting in a time resolution of $\sigma(T_0) = 73.2 \pm 1.2~\ps$ for the trigger system. 

The time difference between the arrival time in the $\bafty$ crystal ($T_{\rm crys}$) and the trigger system was given by $\Delta T = T_0 - T_{\rm crys}$, as shown in Fig.~\ref{CR_test}(b). Fitting to the distribution of $\Delta T$ with a Gaussian function, we obtained  
$\sigma(\Delta T) = 110.6 \pm 1.8~\ps$. Therefore, we obtained the time resolution of the $\bafty$ detector to be  $\sigma(T_{\rm crys}) = 82.2 \pm 2.6~\ps$. In this test, the energy level broadening of the two peaks was minimal,
and background interference with the signal measurements was virtually nonexistent.

\subsection{Electron beam test at SSRF}

\begin{figure}[htbp!]
\centering
\includegraphics[width=0.45\textwidth]{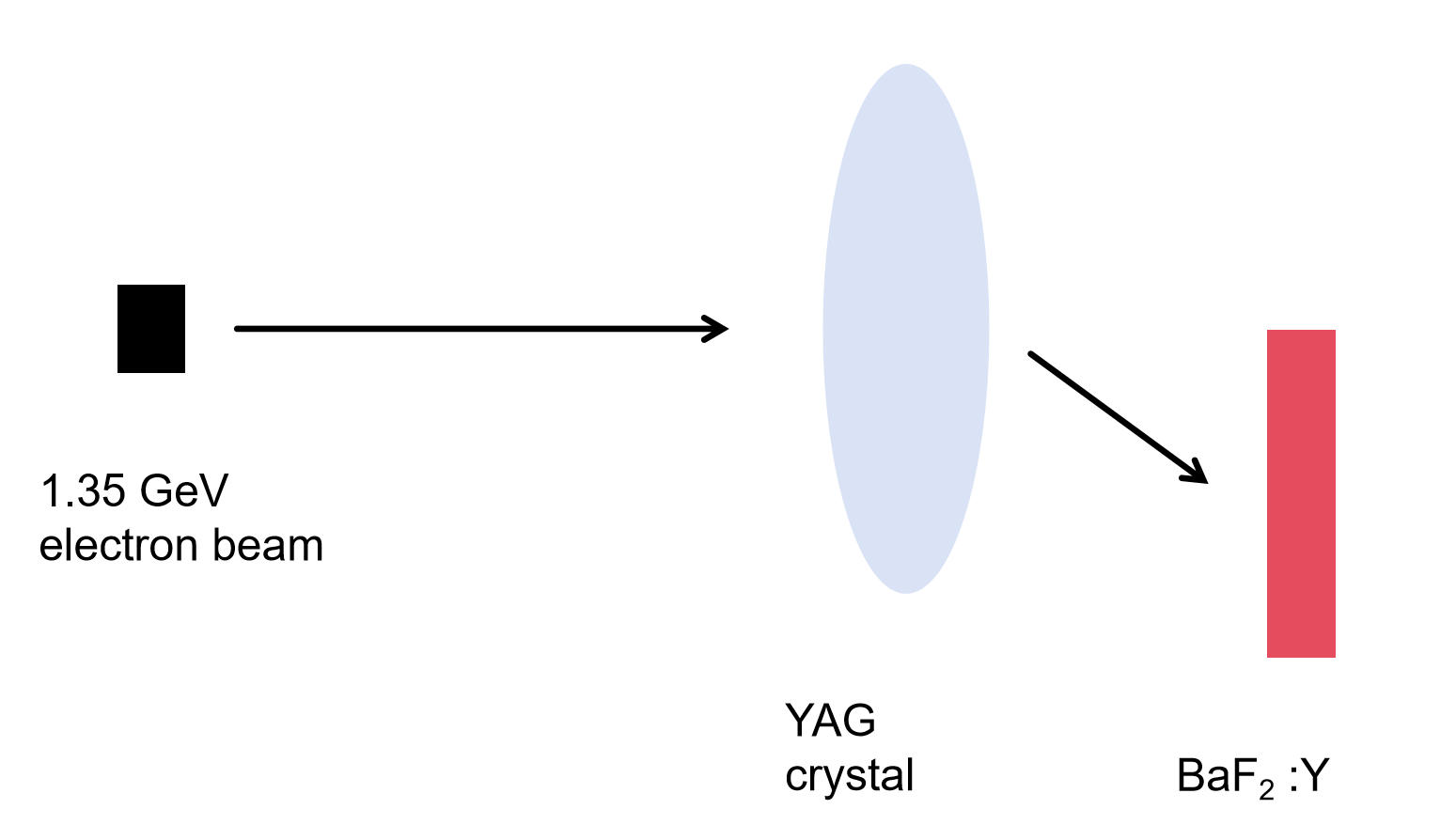}
\caption{The schematic diagram of the setup in the electron beam test at SSRF.}
\label{fig:Detector03}
\end{figure}

To measure the time resolution of the $\bafty$ crystal and its corresponding photon detector for high-energy particles, we positioned the test system within a high-energy electron beam at the SSRF for comprehensive testing. As illustrated in Fig.~\ref{fig:Detector03}, the energy of the electron beam was set to be $1.35~\gev$. Due to the presence of a YAG crystal in the beamline for luminosity monitoring, we could not directly detect the electrons from the beamline but rather those scattered by the YAG crystal. After passing through the YAG crystal, the electrons did not experience significant energy loss, but they had an angular distribution.

During the test, we encountered a significant issue related to the high energy deposited in the two scintillator triggers and the $\bafty$ sample. The pulse heights consistently exceeded the dynamic range of the preamplifier~\cite{pream}, which rendered the CFD method inapplicable. As a result, we relied on timing measurements obtained from a fixed low threshold. \textcolor{red}{A new preamplifier with a higher dynamic range will be designed to fix this issue.}

\begin{figure}[htbp!]
\centering
\includegraphics[width=0.4\textwidth]{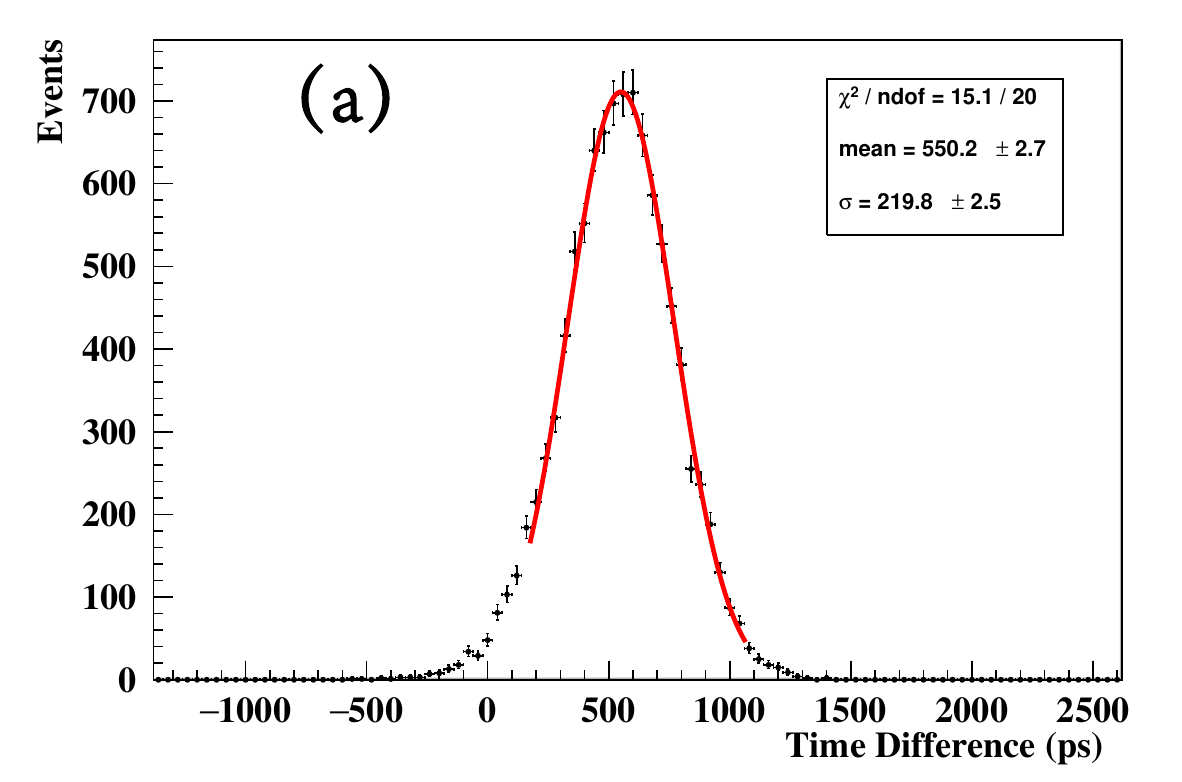}
\includegraphics[width=0.4\textwidth]{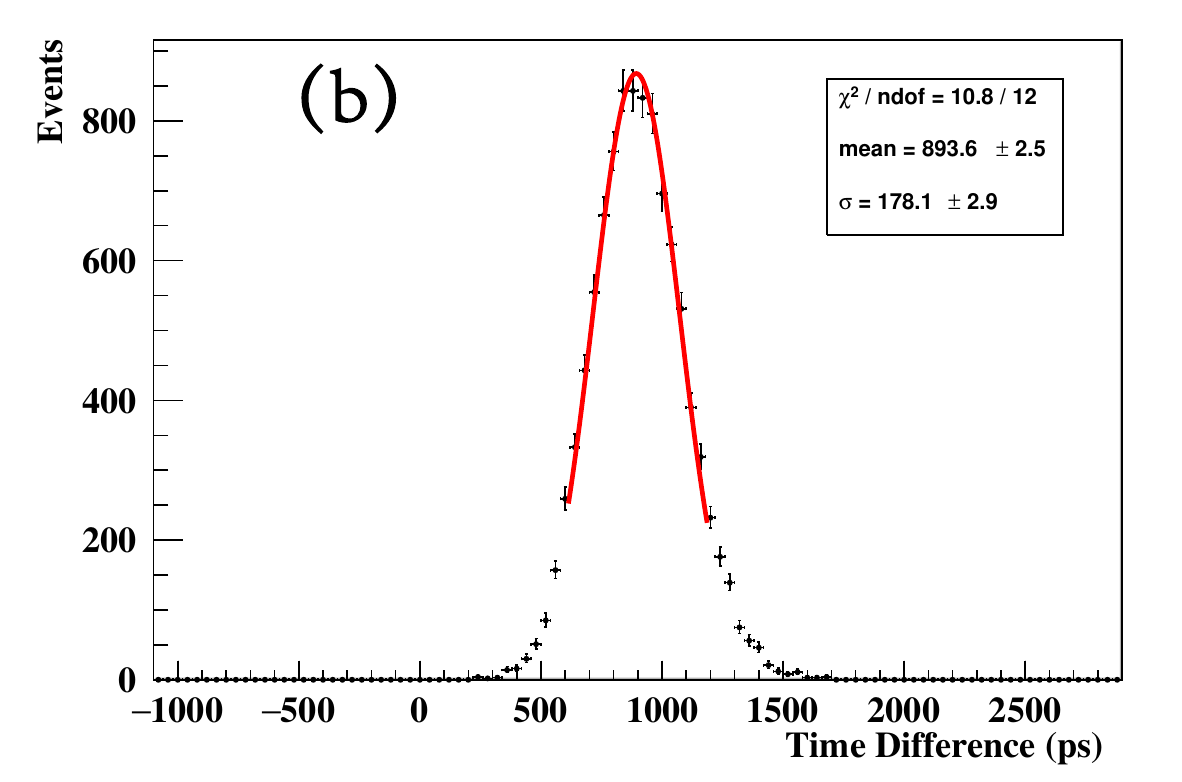}
\caption{Testing with a $1.35~\gev$ electron beam at SSRF. (a) Distribution of the time difference between the two triggers, along with Gaussian fit results. (b) Distribution of the time difference between the $\bafty$ detector and the trigger system, accompanied by Gaussian fit results. }
\label{beam_test}
\end{figure}

By analyzing the $\Delta T_{\rm trg}$ distributions presented in Fig.~\ref{beam_test}(a), we determined a standard deviation of 
$\sigma(\Delta T_{\rm trg}) = 219.8 \pm 2.5~\ps$, which corresponds to a time resolution of $\sigma(T_0) = 109.9 \pm 1.3~\ps$ for this test. The distribution of $\Delta T$ between the $\bafty$ detector and the trigger system is depicted in Fig.~\ref{beam_test}(b), where we obtained $\sigma (\Delta T) = 178.1 \pm 2.9~\ps$, leading to a time resolution of  $\sigma(T_{\rm crys}) = 140.1 \pm 3.8~\ps$.

It is noteworthy that more accurate timing data and energy deposition measurements could be achieved with an enhanced preamplifier that offers a broader dynamic range.

\section{Conclusion and Discussion}

To promote the application of $\baft$ in HEP experiments within particle and nuclear physics, we grew a $200~\mm$-long large $\baft$ crystal doped with 3at\% yttrium. This doping effectively suppresses the slow scintillation component while preserving most of the fast component. XEL measurement results confirmed a consistent reduction in the slow scintillation component throughout the crystal, and ICP-AES indicated a relatively uniform effective yttrium concentration with $K_{\rm eff}=0.75$. The crystal demonstrated an 80\%  lateral transmittance at $200~\nm$ and maintained a transmittance near 90\%  within the visible spectrum, reflecting its excellent optical quality. Additionally, the light response uniformity with a $\delta <3\%$ when coupled with the tail end underscores its suitability for future high-energy experiments applications.

Integrating the $\bafty$ crystal with SiPMs yielded impressive timing resolutions in tests conducted with cosmic rays and high-energy electron beams. Utilizing the CFD method, we achieved a time resolution of $82.2 \pm 2.6~\ps$ for the $\bafty$ detector when exposed to cosmic rays. We could not apply the CFD method in tests with a $1.35~\gev$ electron beam at the SSRF due to excessive deposited energy within the trigger system and the $\bafty$ crystal. Instead, we obtained a time resolution of $140.1 \pm 3.8~\ps$ for the large $\bafty$ detector using timing information from a low fixed threshold.

We anticipate that further optimization of a larger $\bafty$ detector will yield even better time resolution. The promising performance of the $\bafty$ crystal suggests its potential for applications in high-energy experiments. For the next generation of bulk $\bafty$ detectors for HEP applications, we can further optimize the newer and better electronics for the time timing characteristics, also provide further optimization of doping elements, element ratio to improve the light yield of $\bafty$, optimize the growth process to improve the homogeneity of the crystal material to improve the time resolution further.

\section*{Acknowledgment}

This work is partially supported by the National Key R\&D Program of China under Contract Nos. 2022YFA1601903, and
2022YFB3503902; the National Natural Science Foundation of China under Contracts Nos. 12175041, 12025501; the Strategic Priority Research Program of the Chinese Academy of Sciences (XDA25030600); the opening fund of Key Laboratory of Rare Earths, Chinese Academy of Sciences.

\end{document}